\documentclass[twocolumn,preprintnumbers,amsmath,amssymb]{revtex4-1}
\usepackage{graphicx}% Include figure files
\usepackage{ulem}
\usepackage[svgnames]{xcolor}
\usepackage{bm}% bold math
\usepackage{multirow}
\usepackage{multirow}
\usepackage{float}
\usepackage{color}
\begin{document}

\title{Fermi Surface and Carriers Compensation of pyrite-type PtBi$_{2}$ Revealed by Quantum Oscillations}

\author{Lingxiao Zhao$^{1}$, Liangcai Xu$^{1}$, Huakun Zuo$^{1}$, Xuming Wu$^{1}$, Guoying Gao$^{1}$ and Zengwei Zhu$^{1,*}$ }

\affiliation{(1) Wuhan National High Magnetic Field Center\\ School of Physics, Huazhong University of Science and Technology,  Wuhan  430074, China\\
}

\date{\today}

\begin{abstract}
Large non-saturating magnetoresistance has been observed in various materials and electron-hole compensation has been regarded as one of the main mechanisms. Here we present a detailed study of the angle-dependent Shubnikov -de Haas effect on large magnetoresistance material pyrite-type PtBi$_{2}$, which allows us to experimentally reconstruct its Fermi-surface structure and extract the physical properties of each pocket. We find its Fermi surface contains four types of pockets in the Brillouin zone: three ellipsoid-like hole pockets $\alpha$ with C$_4$ symmetry located on the edges (M points), one intricate electron pocket $\beta$ merged from four ellipsoids along [111] located on the corners (R points),  two smooth and cambered octahedrons $\gamma$ (electron) and $\delta$ (hole) on the center ($\Gamma$ point). The deduced carrier densities of electrons and holes from the volume of pockets prove carrier compensation. This compensation at low temperatures is also supported by fitting the field dependence of Hall and magnetoresistance at different temperatures. We conclude that the compensation is the main mechanism for the large non-saturating magnetoresistance in pyrite-type PtBi$_{2}$. We found the hole pockets ¦Á may contribute major mobility because of their light masses and anisotropy to relatively avoid large-angle scattering at low temperature. This may be a common feature of semimetals with large magnetoresistance. The found sub-quadratic magnetoresistance in high field is probably due to field-dependent mobilities, another feature of semimetals under high magnetic fields.
\end{abstract}
\maketitle
%By We carried out angle-depdence of magntoresistnace measurements. The angle-depdence of qauntum oscillations from magnetorestantce mapped the fine structure and volume of the Fermi surface.

Large magnetoresistance and its mechanisms \cite{Mangez,Yntema,Soule,Mun,Takatsu} have drawn tremendous new interest beginning with the discovery of WTe$_{2}$ \cite{Ali2014}. Following this discovery, nonmagnetic materials such as Cd$_{3}$As$_{2}$\cite{Cd3As2MR2014}, WP$_{2}$\cite{WP2MR2015}, LaSb\cite{LaSbMR2015A}, TaAs$_{2}$\cite{TaAs2MR2016}, YSb \cite{Xu2017}, NbP \cite{NbPMR2015}, $\alpha$-As \cite{Zhao2017As}, NbSb$_{2}$ \cite{NbSb2MR2014}, and more recently pyrite-type PtBi$_{2}$\cite{Gao2017PtBi2} have been discovered to show large magnetoresistance. In contrast to the ``conventional" semimetals, bismuth \cite{BismuthMR2009} and graphite \cite{graphiteMR2017}, these materials show non-saturating magnetoresistance as high magnetic field is increased. Several mechanisms have been proposed to explain it. The first mechanism is the electron-hole compensation scenario which has been ascribed to most of the cases listed above, and also other materials\cite{Ali2014,Zhu2015,WP2MR2015,ZengLaSb2016,Guo2016}. Under the e-h ``resonance" condition ($n_{e}=n_{h}$), the magnetoresistivity $\rho(B)\simeq\frac{B^2}{ne}\frac{\mu_e\mu_h}{\mu_e+\mu_h}$ never saturates and exhibits a quadratic dependence in magnetic field, where $n_{e}$($n_{h}$) and $\mu_e$ ($\mu_h$) are the density and mobility of electrons (holes). Other mechanisms range from topological protection \cite{Cd3As2MR2014, LaSbMR2015A, LaSb2016Niu} to metal-insulator transition by magnetic field \cite{graphiteMIT2001,graphiteMIT2001b}.

%While the normal nonmagnetic metals just show a few percent in MR. Materials with XMR have a potential

Recently, pyrite-type PtBi$_{2}$ was predicated a three-dimensional Dirac semimetal \cite{Gibson2015}. Following the prediction, large non saturating magnetoresistance has been experimentally observed in pyrite-type PtBi$_{2}$\cite{Gao2017PtBi2} and its magnetoresistance is in the front rank of the discovered materials, reaching 1.12$\times 10^7$\% at 1.8 K and 33 T. The origin of the large magnetoresistance was preliminarily discussed based on the compensation scenario by fitting with a two-band model \cite{Gao2017PtBi2}. But the detailed Fermi surface and its relation to large magnetoresistance are missing in the previous study\cite{Gao2017PtBi2}. On the other hand, maybe due to the difficulty of peeling the pyrite-structure sample, angle-resolved photoemission spectroscopy (ARPES) data also have not been reported yet. Unlike its hexagonal polymorph PtBi$_{2}$, ARPES results\cite{Thirupathaiah2018, Yao2016} suggest a Dirac-cone-like dispersion may lead to its unconventional large linear magnetoresistance \cite{XuXF2016, YangXJ2016}.

In this paper, we reveal that the compensation of PtBi$_{2}$ would be the mechanism for the large non saturating magnetoresistance by verifying the e-h balance through mapping the  Fermi surface directly by the angle dependence of the Shubnikov -de Haas effect (SdH) and also by a two-band model from temperature-dependence of Hall resistivity and magnetoresistivity. Our mapped Fermi surface from the angle dependence of the SdH suggests four types of Fermi pockets instead of three types of pockets from the previous calculation \cite{Gao2017PtBi2}: three equivalent hole ellipsoid-like pockets named $\alpha$ on the edge($M$ of the Brillouin zone), one electron pocket named $\beta$ located on the corners ($R$), a small octahedron-like electron pocket ($\gamma$) and a large octahedron hole pocket ($\delta$) at the center($\Gamma$) of the Brillouin zone. The electron-hole ratio is compensated within an accuracy of 1\% after summing up the carrier densities from whole obtained pockets by their types. This compensation is also supported by our two-band fitting for the temperature dependence of Hall and magnetoresistance.
%Besides, the mobilities from our SdH and two-band method are consistent, which indicates the XMR could be explained under a classic physics frame.

\begin{figure}
\includegraphics[width=9cm]{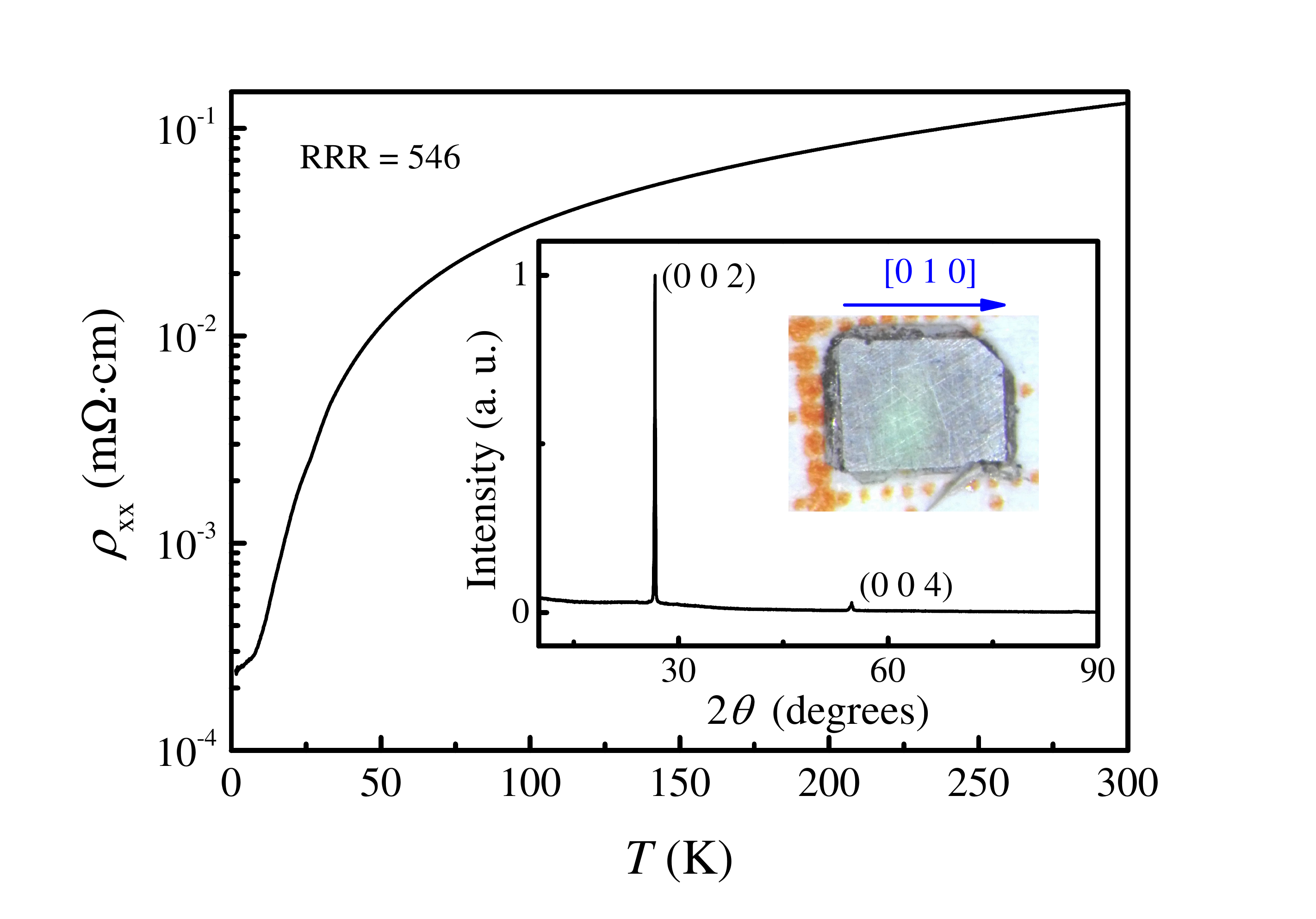}
\caption{ (a) Temperature dependence of resistivity from 1.7 -300 K. The residual resistivity ratio $\rho$(300 K)/$\rho$(1.7 K) is about 546. The inset shows the x0-ray diffraction pattern of the (001) facet of the photographed sample which was then cut into a smaller sample for measurements. The current was along the [010] orientation during whole measurements.}
\end{figure}

Pyrite-type PtBi$_{2}$ single crystals were grown by the flux method with a Pt:Bi = 1:20 molar ratio. Pt powder (99.95\%) and Bi grains (99.999\%) were mixed and sealed in an evacuated quartz ampoule. The ampoule was heated to 450 $^{\circ}$C and preserved for 10 h, then cooled down to 300 $^{\circ}$C in 150 h. After centrifugalizing, single crystals were separated from the flux. The dimension of a typical as-grown sample is around $2\times2\times 2$ mm$^{3}$. The sample was then cut into a narrow strip by a wire-saw to carry on transport measurement. The measurement was performed in a TeslatronPT (Oxford Instruments) equipped with a home made rotator whose angle can be controlled by a data-collecting computer. Angles of the motor-driven rotator were determined either by a Hall probe on the sample holder or by steps from the driving motor which was calibrated before measurement. The two angle-determining methods show good consistency. The electrical current was applied by a Keithley 6221 and the voltage was measured by a Keithley 2182A. The magnetic field was perpendicular to the current during rotation. The high-magnetic-field magnetoresistance measurement was performed in the Wuhan National High Magnetic Field Center. A 100 kHz ac current was applied by a NI-5402 signal generator and voltage was recorded by a NI-5105 high-speed digitizer which worked in 4 MHz. A digital phase lock-in method was applied to extract magnetoresistance.

Figure 1 shows the temperature dependence of resistivity $\rho_{0}(T)$ of the measured sample which has a residual resistivity ratio RRR = $\rho$(300 K)/$\rho$(1.7 K) of 546. More samples with different RRR were measured and show same quantum-oscillation results. Both resistivities at room temperature of 132 and 0.24 $\mu \Omega$ cm at 1.7 K are higher than those of reported \cite{Gao2017PtBi2}. Such differences may be due to the different sample quality or the different orientations of the applied current. Note that the current was injected along [0\={1}1] in the previous report \cite{Gao2017PtBi2} while it was along [100] in our case, which could induce difference in carrier mobility. For instance, the mobility is closely related to the orientation of the applied current in bismuth\cite{Collaudin2015}. The inset of Fig. 1 shows the x-ray diffraction pattern of the photographed sample whose crystallographic direction [010] is labeled with a blue arrow. The sharpness of the (002) indicates the high quality of the sample.

\begin{figure}[!htb]
\includegraphics[width=9cm]{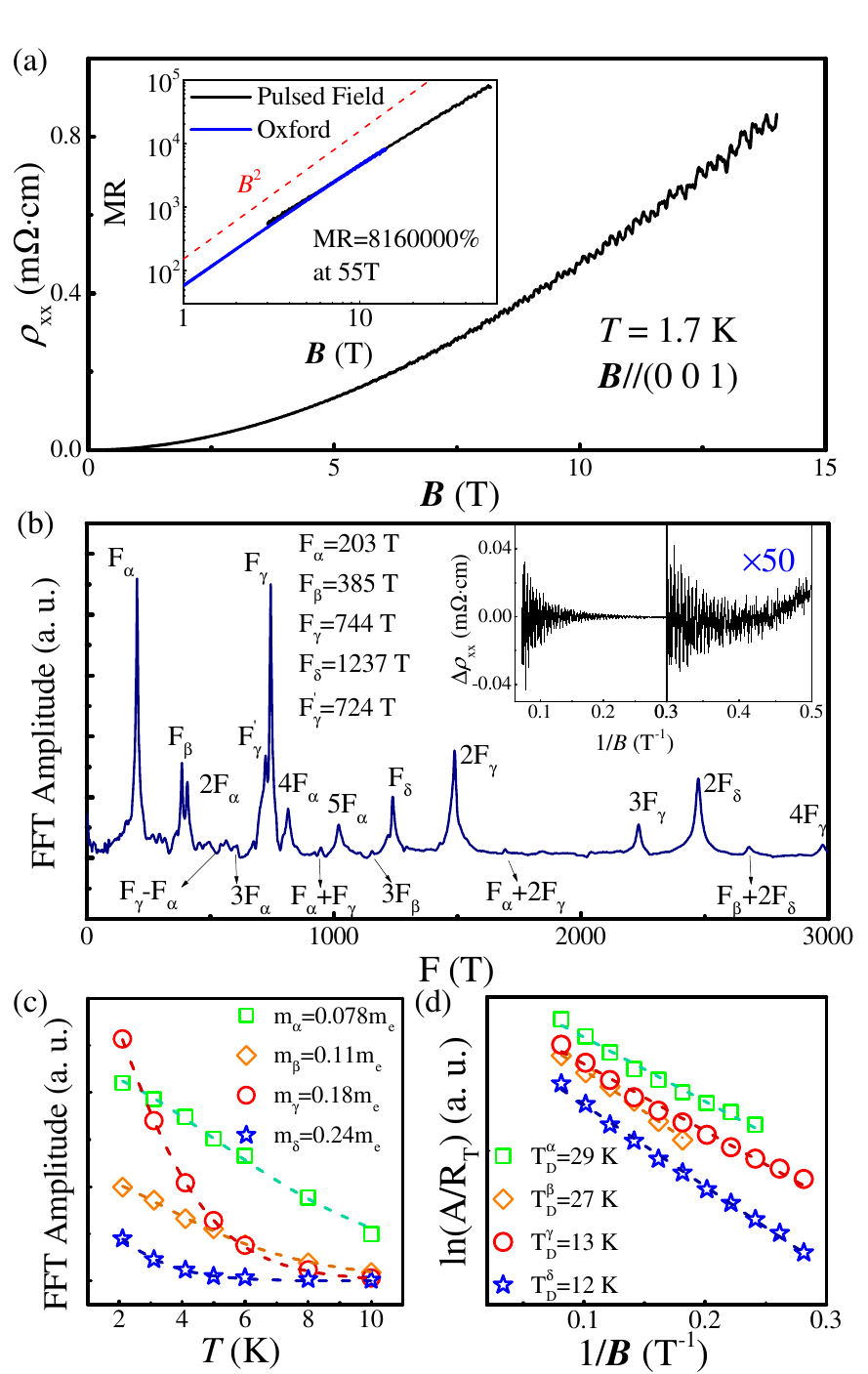}
\caption{ (a) Field-dependent $\rho (B)$ curve at 1.7 K as magnetic field along [001] and the current along [100] using a TeslatronPT from Oxford Instruments. The inset shows the MR ($[\rho(B)-\rho(0)]/\rho(0)\times100\%$) from the same data in the  blue curve and also with data from pulsed fields up to 55 T where the MR reaches 8 160 000\%. The MR curve has a subquadratic dependence in magnetic fields while a red-dashed $B^2$ curve is offset for comparison.  b). FFT spectra of SdH oscillations after subtracting the background from magnetoresistance. The SdH oscillations as a function of 1/ $B$ are also shown in the inset. The oscillations start at as low as 2 T indicating high mobility of the sample. We also indexed the sharp peaks of the FFT spectra accordingly. (c), (d) The cyclotron masses and Dingle temperatures of different pockets as field along [001] extracted from the temperature dependence of the SdH effect with the Lifshitz-Kosevich theorem \cite{QuantumOscillaitons}.}
\end{figure}

\begin{figure*}[!htb]
\includegraphics[width=17cm]{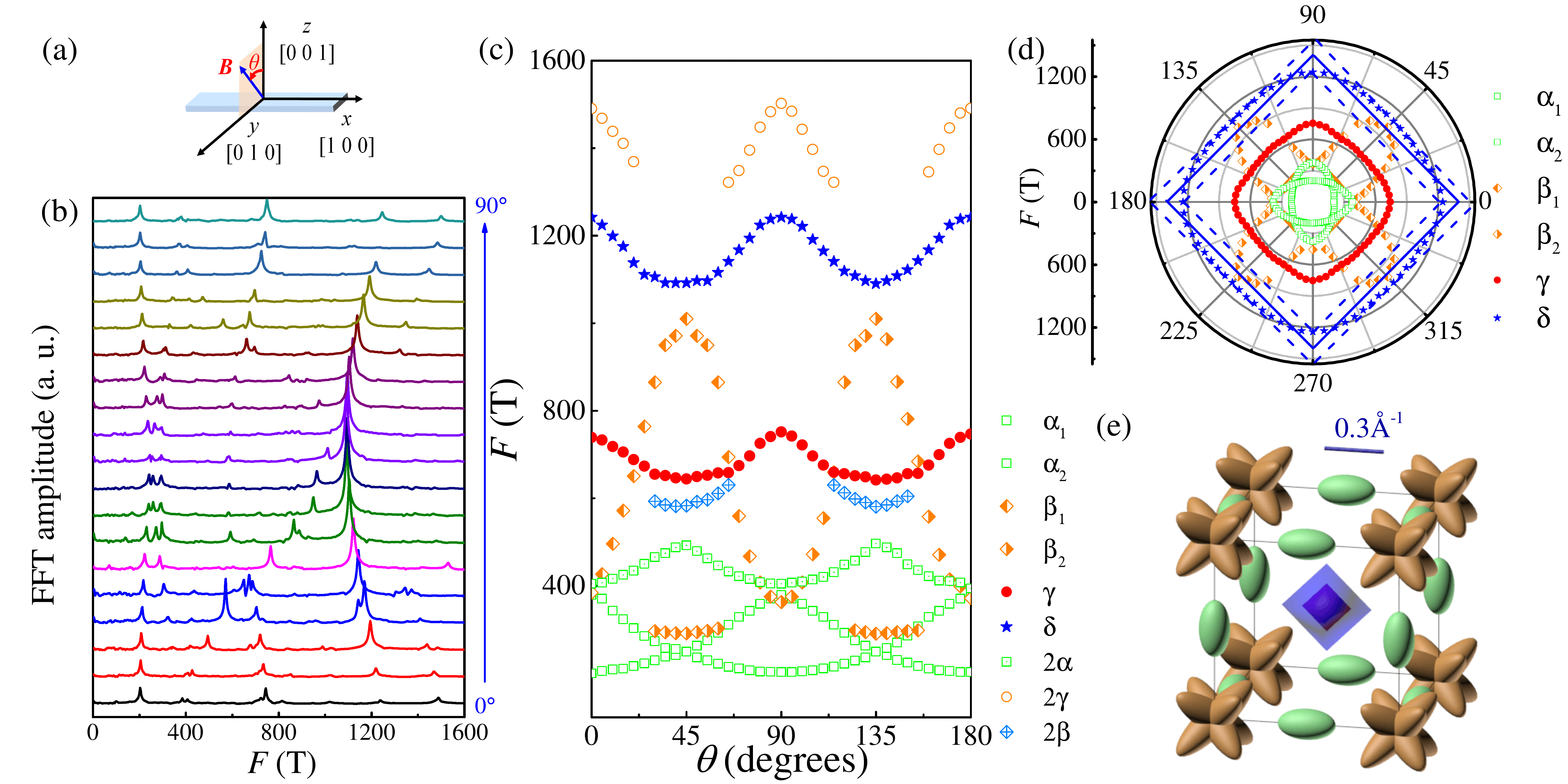}
\caption{ (a) Schematic diagram of the sample geometry for measurement. The magnetic field was rotated in the [001]-[010] plane as the current along [100]. (b) Shifted FFT spectra of SdH as a function of frequency every 5$^\circ$  from 0 to 90$^\circ$. (c) Angular dependence of SdH frequencies. The symbols are for experimental data from the peaks of the FFT spectra. We can identify four fundamental frequencies and their harmonic frequencies. (d) The fundamental frequencies of four pockets in a polar plot where two octahedron-like Fermi pockets ($\delta$ and $\gamma$) can be easily identified, besides ellipsoid-like pockets $\alpha$ and $\beta$. In the fitting of the octahedron, we take the geometric mean (in the solid line) of two minimal and maximal edge lengths shown in dashed lines for the length of its edge. (e) Fermi-surface reconstruction from the fitted parameters. The locations of pockets are determined from symmetric and size arguments (see the main text). A scale bar is also shown. }
\end{figure*}

Figure 2(a) displays a typical field-dependent $\rho(B)$ curve with pronounced SdH oscillations at 1.7 K with the magnetic field along [001] and the current along [100]. The inset shows the magnetoresistance (MR) ($[\rho(B)-\rho(0)]/\rho(0)\times100\%$) $\sim$ 817 300\% at 14 T from the same data and also shows MR up to $\sim$ 8 160 000\% at 55 T without any signature of saturation measured in a pulsed magnet. These results confirm the large non-saturating magnetoresistance in pyrite-type PtBi$_{2}$, even the RRR is lower than that reported in \cite{Gao2017PtBi2}. The MR curve can be fitted with a power law MR $=aB^{1.79\pm0.02}$, deviating from a quadratic dependence in field. The inset of Fig. 2(b) presents the SdH oscillations after subtracting the background by a polynomial fitting. The SdH oscillations can be clearly seen from $B$ = 2 T, indicating high carrier mobility of the sample by a first approximation to have the mobility of $\mu\geq$ 0.5 T $^{-1}$. These oscillations allow us to extract the fine structure of the Fermi surface. The SdH oscillations as a function of 1/ $B$ are also shown in the inset, these oscillations are observable down to 0.5 T$^{-1}$ after enlarging 50 times. We then obtain fast Fourier transform (FFT) spectra of SdH oscillations and plot them in Fig. 2(b). Each FFT peak at the fundamental or higher harmonics can calibrate the extremal Fermi surface area which is perpendicular to the magnetic field \cite{QuantumOscillaitons}. We can clearly identify and index the sharp peaks of the FFT spectra for four different frequencies named $\alpha$, $\beta$, $\gamma$, and $\delta$ in Fig. 2(b).  Note that the amplitude of the second-harmonic frequency of $\delta$ is higher than that of the fundamental frequency. This is probably due to spin splitting of this band \cite{QuantumOscillaitons} and confirmed in our later angle-dependence of the SdH effect: the amplitude of the fundamental frequency of $\delta$ becomes larger as a normal oscillation series does after the rotation. The small amplitude of 3$F_{\alpha}$ is also due to spin splitting \cite{SM}. The cyclotron masses and Dingle temperatures for four pockets as field along [001] are shown in Figs. 2(c) and (d), extracted from the SdH curves at different temperatures with the Lifshitz-Kosevich formula \cite{QuantumOscillaitons} by the attenuation factors due to finite temperature: $R_T=\frac{X}{\rm {sinh} X}$ and impurity scattering $R_D= \rm{exp}(-\frac{\pi m^*}{eB\tau_D})$ where $X=\frac{2\pi^2k_BTm^*}{e\hbar B}$. We note that our results are not exactly the same as those in Ref. 15, where a different orientation of magnetic field is applied along [111]. We tried the de Haas -van Alphen (dHvA) effect and find similar masses from the SdH\cite{SM} in our case. We also notice that the theoretical results proposed three pockets in Ref. 15, instead of the four pockets found in the current work. Fine-tuning of the parameters is probably needed for the theoretical calculation since the spin-orbit coupling and electron correction should play important roles in the compound.

Now we can map the Fermi surface by rotating the sample to the get angle dependence of SdH; the schematic diagram of the sample geometry is shown in Fig. 3(a). Through the above procedure, we obtained the FFT spectra of SdH at various angles by rotating the sample at 1.7 K from [001] to [00\={1}]. Figure 3(b) shows the shifted FFT spectra every 5$^\circ$ from 0 [001] to 90$^\circ$ [010]. We have not shown the data beyond 90$^\circ$, since the angle dependence of the SdH pattern exactly repeats because of its cubic crystal structure nature. Figure 3(c) shows the angle-dependence of the FFT peaks in different symbols. The branch $F_{\alpha_1}(\theta)$ displaces 90$^\circ$ from the branch $F_{\alpha_2}(\theta)$. So we deduce that the three equivalent $\alpha$ pockets locate on the edge because of their C$_4$ symmetry with an assumption that pockets always have the highest symmetric locations. This assumption is also for determining locations of other types of pockets. Another almost constant frequency $F_{\alpha_3}(\theta)$ spectrum is absent in the current results. This absence should be due to the lower mobility when the current is along the long axis of the $\alpha_3$ ellipsoid, as seen in YSb\cite{Xu2017}. Then for the $\alpha$ pockets which are prolate spheroids, we fitted the angle dependence of $F_{\alpha_1}$ quantitatively by the following equation:
\begin{small}
\begin{equation}\label{s}
F_{\alpha_i}=F_{0}/\sqrt{(\cos[\theta-(i-1)\pi/2])^2+(\lambda\sin[\theta-(i-1)\pi/2])^2}
\end{equation}
\end{small}
Where we can obtain $F_{0}$ = 202 T, $\lambda=0.55$ and ``$i$" is for the subscript of $\alpha$. By Onsager relation $F=(\hbar/2\pi e)A_k$ between frequency $F$ and the extreme cross section $A_k$ of a Fermi surface, we extracted the values listed in Table \ref{tableVolume}. The carrier density for each equivalent $\alpha$ prolate spheroid is $0.29\times10^{20}$ cm$^{-3}$. The type of this pocket is hole, by using the previous calculations \cite{Gao2017PtBi2} as a guide. The types of other pockets are deduced by the same route.

The $\beta$ pocket has quite a0 complicated structure, but can be sorted out. We first exclude that these ellipsoids exist independently, locating between $\Gamma$ and $R$ along [111]. The shape from $F_{\beta}(\theta)$ resembles that of an ellipsoid, but is rotated 45$^\circ$ along the [100] axis at first glimpse. According to the highest symmetry requirement, the long axis of the ellipsoid should lie at [111] to have a maximum cross section of an ellipsoid at 45$^\circ$ in the current case. Note that this ellipsoid is then tilted when the field is rotated, as the rotating axis is not along its major or minor axis. Then Eq. (\ref{s}) is no longer valid for this case. By geometric consideration of an independent ellipsoid, we obtain its semi-minor axis $k_{a}$ = $k_{b}$ = 0.094 {\AA}$^{-1}$ and its semi-major axis $k_{c}$ =  0.33 {\AA}$^{-1}$ from the FFT values at high symmetric angles. 7.	But the $\beta$ band will touch two other bands $\gamma$ and $\delta$ (discussed below) by considering that two semimajor axes of ¦Â bands lie along [111] according to the symmetry, note that the diagonal length of the Brillouin zone is only 1.62 {\AA}$^{-1}$. More quantum oscillations are expected from this touching, which contradicts our observations. Besides, the carrier density from eight such pockets would surpass 10$^{21}$ cm$^{-3}$, which contrasts with the semimetal property from its resistivity. Therefore, these $\beta$ pockets should depend on and have to cross each other, leading to a reduction of the total number of ellipsoid also its length [illustrated in Fig.3(d)]. We obtain the volume of this pocket through summing up the volume of four ellipsoids whose middle parts are truncated and one middle sphere. By considering the frequencies of an ellipsoid tilted along [111] when the field is at 0, 45$^\circ$ and also the fact that the electrons actually travel across two connected ellipsoids at $\theta=0$ in this case, we extract actual $k_{a}$ = $k_{b}$ = 0.0792 {\AA}$^{-1}$ and $k_{c}$ =  0.1584 {\AA}$^{-1}$ and the radius of the middle sphere is $k_{s}$ = 0.1094 {\AA} $^{-1}$ from the lower frequency of $F_{\beta}$ at 45$^\circ$. Finally, the carrier density of $1.88\times10^{20}$ cm$^{-3}$ for this electronlike pocket is obtained.

Although the shape is not an exact square in the polar plot Fig. 3(d) as for an octahedron, we can still treat the $\gamma$ and $\delta$ bands as two smooth and cambered octahedrons whose symmetry naturally meet the requirement of a cubic. This cambered surface induces a warping effect to have additional $F_{\gamma}^{'}$ which have the same angle-dependent behavior as $F_{\gamma}$\cite{SM}, but vanishes at larger angles. These octahedrons should locate at the corner or the middle of the Brillouin zone because of their high symmetry. Under this assumption, we calculated the geometric mean of the edge length [the solid line in Fig. 3(d)] of the minimal and maximal octahedron of $\delta$ which are able to encircle the band [the dash lines in the fig. 3(d)]. The same procedure is used to obtain the edge length of octahedron for $\gamma$. After calculating volumes, we deduce the densities of electron-like $\gamma$ and hole-like $\delta$ are 0.84$\times10^{20}$ and 1.82$\times10^{20}$ cm$^{-3}$, respectively.

\begin{table*}[!htb]
\begin{tabular}{| p{3cm} | p{3cm} | p{3cm} | p{3cm} | p{3cm} |}
\hline
 & $\alpha$ ($h$) &  $\beta$($e$) & $\gamma$ ($e$) & $\delta$ ($h$)  \\
 \hline
Location & Edge (M) &  corner (R)& Center ($\Gamma$) & Center ($\Gamma$)\\
\hline
\multirow{6}{*}{\textit{k} ({\AA}$^{-1}$)} &  & mid. sphere& &  \\
  &  & $k_s$=0.1094  &  &  \\
 & $k_{a}$=0.0777  & ellipsoids & $k_{edge}$=0.198 & $k_{edge}$=0.257    \\
  &  $k_{b}$=0.0777   & $k_{a}$=0.0792 &  &    \\
    &  $k_{c}$=0.143    & $k_{b}$=0.0792 &  &    \\
   & & $k_{c}$=0.1584 &  &   \\
\hline
 $V$({\AA}$^{-3}$)  & 0.0036&0.0221 &0.0104  &0.0225   \\
\hline
 Number  & 3&1 &1  &1   \\
\hline
\textit{n}(cm$^{-3}$)&$0.87\times10^{20} $ & $1.88\times10^{20} $ &$0.84\times10^{20}$ & $1.82\times10^{20}$ \\
\hline
fraction&1.32\% & 2.68\% &1.26\% & 2.73\% \\
\hline
$m_{cyc.}$(m$_e$) & 0.078$\pm0.001$&0.11$\pm0.01$&0.18$\pm0.01$&0.24$\pm0.01$\\
\hline
$T_{D}$ (K)&29&27&13&12\\
\hline
$\tau_{D}$ (s)&4.2$\times10^{-14}$&4.5$\times10^{-14}$&9$\times10^{-14}$&1$\times10^{-13}$\\
\hline
$\mu_{D}$ (cm$^2$/Vs)&944&719&913&742\\
\hline

\end{tabular}
\caption{Summary of physical properties: volume, quantity, carrier density (\textit{n}), cyclotron mass ($m_{cyc.}$), dingle temperature $T_{D}$, relaxation time $\tau_{D}$ and mobility ($\mu_{D}$) of $\alpha$, $\beta$, $\gamma$ and $\delta$ pockets. The momentum [$k_{a}$, $k_{b}$ and $k_{c}$], $k_{edge}$ and $k_{s}$ were axes of an ellipsoid, an edge length of the octahedron Fermi surface, and a radius of the sphere in the middle of $\beta$, respectively.}
\label{tableVolume}
\end{table*}

Fig. 3(e) is a scale drawing in a certain scale(a scale bar is shown), in which we summarized and reconstructed the Fermi surface according to our SdH results. Note that the symmetry of $\beta$, $\gamma$ and $\delta$ are same. So another possibility of locations of pockets is that $\gamma$ and $\delta$ locate on the corners and $\beta$ is on the center of the Brillouin zone. However, the total hole ($\alpha$ and $\delta$) and electron ($\beta$ and $\gamma$) carrier densities not affected in the two cases are $2.69\times10^{20}$ and $2.73\times10^{20}$cm$^{-3}$, respectively. This result suggests compensation between electrons and holes at a ratio of 0.99. The carrier densities are typical for semimetals, such as WTe$_{2}$($6.6\times10^{19}$) \cite{Zhu2015} ,LaSb($1.1\times10^{20}$) \cite{LaSbMR2015A}, Sb($5.5\times10^{19}$) \cite{SbDensity}, WP$_{2}$($1.4\times10^{20}$) \cite{WP2MR2015}, and also $\alpha$-As($1.1\times10^{20}$)\cite{Zhao2017As}. And all their mobilities are around $10^{5}$ (up to $10^{6}$ in Sb) cm$^{2}$ V$^{-1}$ s$^{-1}$. Such shared common properties may be the key to understand large magnetoresistance \cite{ShallowFS}. For a compensated semimetal, we would expect a quadratic dependence of magnetoresistance, but the MR=$aB^{1.79\pm0.02}$ in this material. Such deviation of the quadratic dependence of magnetoresistance is prevalent, due to an unavoidable field-induced reduction in mobility by disorders \cite{Sb2018}.

To further illustrate the compensation, we carried out temperature dependence of magnetoresistance and Hall resistivity, shown in Figs. 4(a) and (b), and found a similar result as in the previous report\cite{Gao2017PtBi2}. We extracted the carrier densities and mobilities by fitting the Hall conductivity by the two-band model\cite{He2016,Pavlosiuk2016,Zhao2017As}:

$\sigma_{xy}(B)=\frac{\rho_{yx}(B)}{\rho_{xx}^2(B)+\rho_{yx}^2(B)}=[\frac{n_{h}\mu_{h}^2}{1+(\mu_{h}B)^2}-\frac{n_{e}\mu_{e}^2}{1+(\mu_eB)^2}]eB$ ,

present in Figs. 4(c) and (d). Although there is a slight discrepancy between electron and holes at the high-temperature range, the carrier density between two types of carriers becomes equal below 20 K and the ratio $n_e/n_h$ is about 0.99 at 1.7 K from fitting. The carrier densities $n_e=1.32\times10^{20} $cm$^{-3}$ and $n_h=1.33\times10^{20}$ cm$^{-3}$ are also quite close to the values from the Fermi-surface method. This two-band model gives relatively lower numbers which should be higher since the carrier density easily surpasses $1.8\times 10^{20}$ cm$^{-3}$ even only from the $\delta$ band. Two other two-band fitting methods from magneto-conductivity and Hall resistivity shown in the Supplemental Material \cite{SM} again exhibit compensation of carriers, but get higher carrier densities. The difference of carriers density between different fitting methods indicates the limitation of two-band fitting for a multiband system.

%The two-band model usually gives lower carrier densities\cite{}.
The mobility of holes increases faster than that of electrons to $1.5\times 10^{5}$ cm$^{2}$ /Vs  as the temperature is lowered. This increase of hole mobility may mainly result from the contribution of $\alpha$ pockets which have the lightest masses and their anisotropy($k_c/k_a\simeq2$). Such two features are prevalent in the semimetals at least in one of their Fermi pockets: Sb \cite{Sb2018, SbDensity}($m^*\sim0.088m_e$, anisotropy of $k \sim5.2$) , bismuth \cite{bismuthF} ($m^*\sim0.0011m_e$, anisotropy of $k \sim15$), YSb \cite{Xu2017,YSbanistropy,YSbmass}($m^*\sim0.2m_e$, anisotropy of $k \sim2$) and WTe$_2$ \cite{Zhu2015,WTe2mass}($m^*\sim0.4m_e$, anisotropy of $k \sim2-3$). The $\frac{\mu_{T}}{\mu_{D}}$ is around 150 and 25 for holes and electrons, respectively, and $\mu_{T}$ is the mobility from the two-band fitting. This indicates that a small-angle-scattering process in quantum oscillation plays a significant role during carrier transport after electron-phonon scattering has faded as the temperature is decreased, which may be a common feature in semimetals \cite{Sb2018}. The quantum oscillation mobilities are integral to all the scattering angles while the mobilities of Hall effect are only affected by large-angle scattering \cite{Gao2017PtBi2,Sb2018}.
So, to have a pocket with light mass and anisotropy may relatively avoid large-angle scattering at low temperature to have higher transport mobility.
% The mobilities found here are similar to these found from our SdH measurement. So this material may be lack of the topological protection which was suggested in previous report.

\begin{figure}
\includegraphics[width=9.5cm]{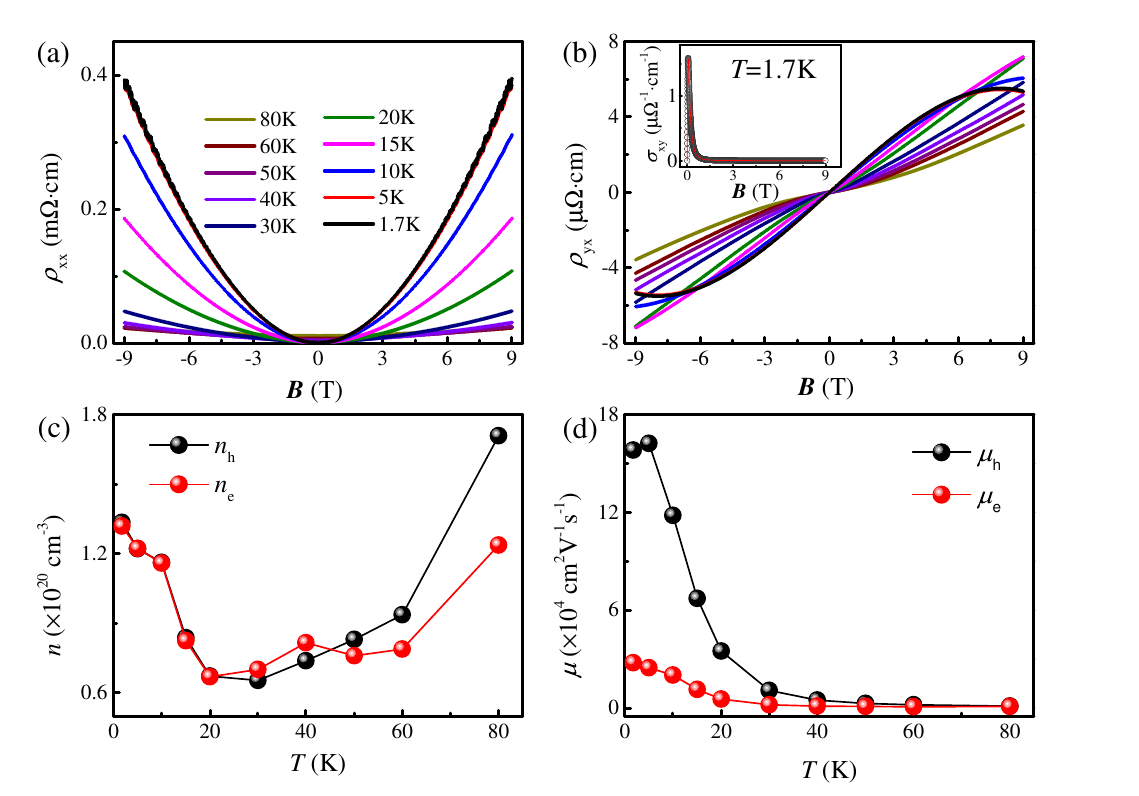}
\caption{ a)Temperature-dependence of magnetoresistance and b) Hall resistivity in different temperatures. The inset shows a red fitting line from two-band model and black $\sigma_xy$ symbols measured at 1.7 K . c, d). Temperature-dependence of carrier density $n_{e,h}$ and  $\mu_{e,h}$. The two-band model again supports the compensation of electron and hole from the Fermi-surface method.  }
\end{figure}

In summary, we have experimentally mapped out the Fermi surface of pyrite-type PtBi$_{2}$ including two hole and two electron pockets by angle-dependence of SdH measurements. This allows us to deduce the carrier density and mobility of each pocket and to reveal a compensation between the electron and hole. Such compensation is further supported by two-band fitting from the temperature dependence of magnetoresistance and Hall resistivity. We ascribed the large non saturating magnetoresistance found in this material to the compensation. The high hole mobility may be due to the light mass and anisotropy of $\alpha$ pockets.

We acknowledge useful discussions with Kamran Behnia, Beno\^{\i}t Fauqu\'{e} and Gang Xu. This work is supported by the 1000 Youth Talents Plan, the National Science Foundation of China (Grants No. 11574097 and No. 51861135104), the National Key Research and Development Program of China (Grant No.2016YFA0401704) and the China Postdoctoral Science Foundation (2018M630846).

%We now turn our attention to the amplitude of quantum oscillations. During the field-rotation, we notice the amplitude of FFT spectrum of Fd as the field along [011] is up to 6 times larger than that of a field along [001], while the magnetoresisitivty are small for both cases. Such amplitude enhancement could be from two sources dramatic enhancement of effective mass or spin mass following the full Lifshitz-Kosevich(LK) equation for oscillatory part:
%\begin{small}
%\begin{equation}\label{s}
%\frac{\Delta\rho}{\rho_{0}}= AR_{T}R_{D}R_{S}cos[(2\pi[\frac{F}{B}-\frac{1}{2}+\phi_{D}])
%\end{equation}
%\end{small}
%Where $R_{D}=exp(-\pi\sqrt{\frac{2\hbar F}{e}}\frac{1}{\ell B})$ and $R_{S}=cos(\frac{\pi glm_{s}}{2m_{0}})$ are the impurity and spin damping factors respectively. At the same tempeorature, the oscillatory amplitude is determined by R$_{T}$(m) and R$_{s}$(m$_{s}$). In order to calify this, we carrier out to temperature-dependence of magnetresistance as the field along [011] to extract the mass of four pockets. The effective mass of $\delta$ band is 0.26 m$_{e}$ which is exact same as that in [001] case. So we conclude that the spin mass of $\delta$ is drastically enhance when the field is along [011]. To further illustrate this spin-mass enhancement, we extracted the angle-dependent amplitude of FFT spectra. Remarkable, the amplitude is following a cosine shape which is mark of 2D system. In 2D system, the spin-mass is always assumed to have cosine dependence $m_{s}(\theta)=\frac{m_{s}}{cos\theta}$

\noindent
* \verb|zengwei.zhu@hust.edu.cn|\\

\clearpage
%\appendix       %%% starting appendix
%\section{Supplementary Material for: Graphite in 90 T: Evidence for Strong-coupling Excitonic Pairing}

%\author{Zengwei Zhu$^{1,2,*}$, Pan Nie$^{1}$, Beno\^{\i}t Fauqu\'{e}$^{3,4}$, Ross D. McDonald$^{2}$, Neil Harrison$^{2}$ and Kamran Behnia$^{1,4}$}

%\affiliation{(1) Wuhan National High Magnetic Field Center and School of Physics, Huazhong University of Science and Technology,  Wuhan  430074, China.\\
%(2) MS-E536, NHMFL, Los Alamos National Laboratory, Los Alamos, New Mexico, 87545, USA.\\
%(3) JEIP,  USR 3573 CNRS, Coll\`ege de France, PSL Research University, 11, place Marcelin Berthelot, 75231 Paris Cedex 05, France.\\
%(4)Laboratoire de Physique Et Etude des Mat\'{e}riaux (UPMC-CNRS), ESPCI Paris, PSL Research University 75005 Paris, France.\\}
%
%%\date{\today}
%\maketitle
{\large\bf Supplemental Material for ``Fermi Surface and Carriers Compensation of pyrite-type PtBi$_{2}$ Revealed by Quantum Oscillations''}

\setcounter{figure}{0}
\setcounter{equation}{0}
\setcounter{table}{0}

\section{The effective masses from dHvA}
We studied the de Haas¨Cvan Alphen (dHvA) oscillations of our sample with magnetic field B parallel to [001]. The inset of Fig. S1 (a) presents the dHvA oscillations after subtracted the background by fitting polynomial fitting for $T$ = 1.8 K. Then, fast Fourier transform (FFT) were done and plotted in Fig. S1 (a). We can clearly identify three set of fundamental frequencies labeled F$_\alpha$, F$_\beta$, and F$_\gamma$, which are corresponding to the result from SdH oscillations. The cyclotron masses of different pockets were extracted from temperature-dependence of dHvA with Lifshits-Kosevich theorem shown in Fig. S1. (b). The frequencies and cyclotron masses of F$_\alpha$, F$_\beta$ and F$_\gamma$ from SdH oscillations and dHvA oscillations are very close. But due to the fast scan and limit of magnetic field, the amplitude of the F¦Ä is too weak to determine its effective mass. The cyclotron masses of three pockets from dHvA oscillations are consistent with that from SdH.
\begin{figure}[!htb]
\includegraphics[width=9cm]{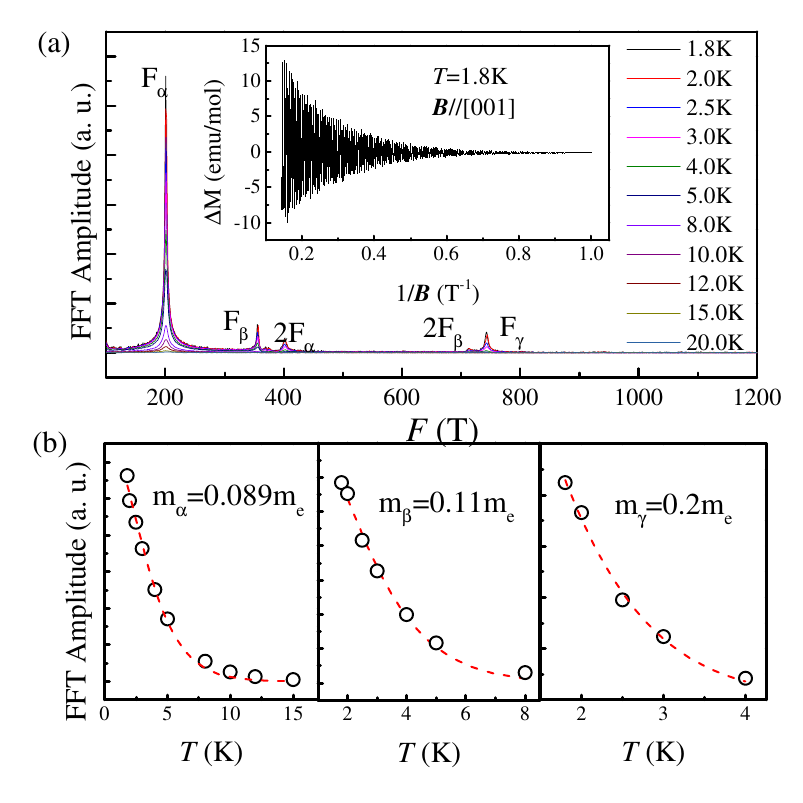}
\caption{ (a) The FFT spectra of dHvA oscillations with magnetic fields along [001] for PtBi$_{2}$ at various temperatures. The dHvA oscillations as a function of 1/B are also shown in the inset at 1.8 K. We indexed the sharp peaks of FFT spectra accordingly. (b) The cyclotron masses of different pockets as field along [001] extracted from the temperature dependence of dHvA with the Lifshits-Kosevich theorem.}
\end{figure}

\section{The magnetoresistance up to 80 T}
The MR reaches 15,500,000\% in pulsed fields up to 80 T (black curve), exhibiting a sub-quadratic dependence of magnetic fields. The orange-dashed $B^2$ line is offset for comparison.
\begin{figure}
\includegraphics[width=9cm]{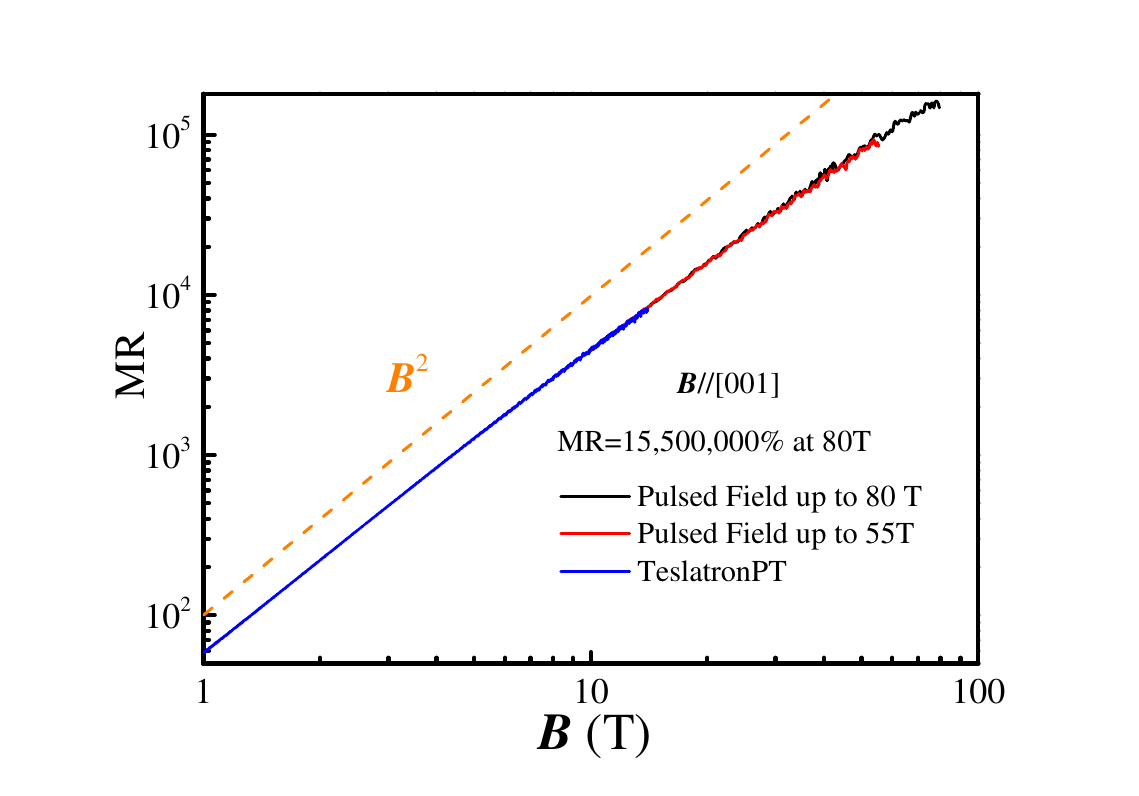}
\caption{ The MR curves collected at \textit{T} = 1.7 K with magnetic field parallel to the [001] direction and current direction along [100] by using Teslatron from Oxford Instrument and pulsed field. The MR reaches 15 500 000\% in pulsed fields up to 80 T (black curve), exhibiting a sub-quadratic dependence of magnetic fields. The orange-dashed $B^2$ line is offset for comparison.}
\end{figure}
\section{The quantum oscillations from Hall resistivity}

Fig. S3. shows the oscillations part of Hall resistivity at 1.7 K. We can clearly see the SdH oscillations can be clearly seen in the Fig. S3. (a), and four sets of fundamental frequencies F$_\alpha$, F$_\beta$, F$_\gamma$ and F$_\delta$  which are of the same value obtained from the magnetoresistivity, can be distinguished in the Fig. S3 (b).
\begin{figure}
\includegraphics[width=9cm]{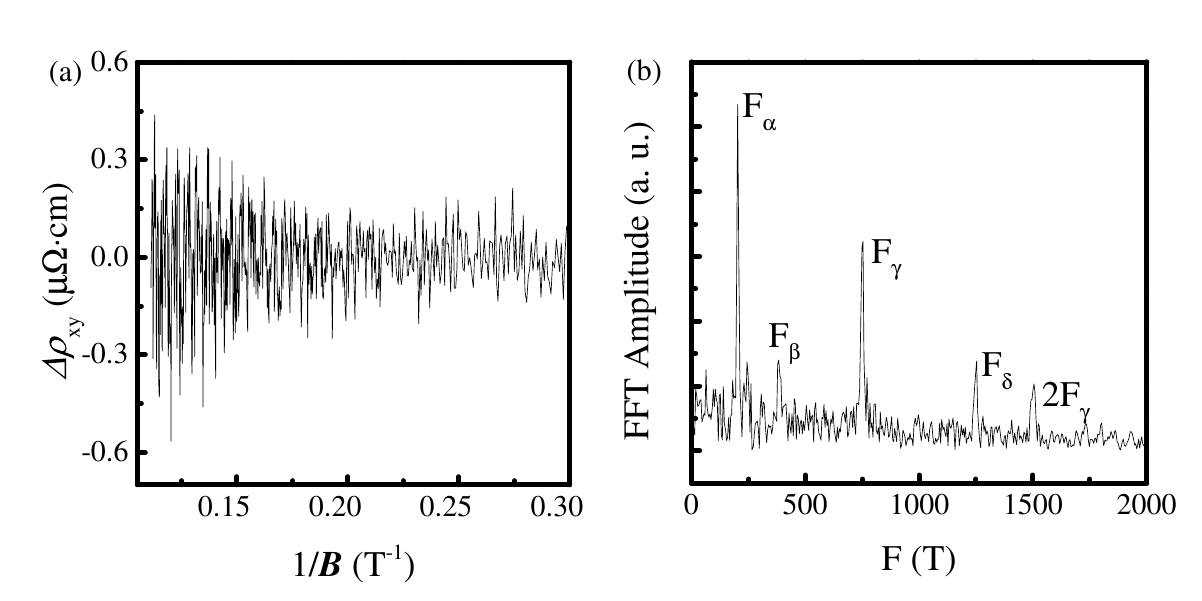}
\caption{ (a) The SdH oscillation part of Hall resistivity ($\Delta\rho_{xy}$) plotted as a function of inverse field at 1.7 K. (b) FFT spectra of SdH oscillations from Hall resistivity.}
\end{figure}

\section{The amplitude of 3F$_\alpha$}
The amplitude of 3F$_\alpha$ can be small by considering spin-splitting due to Zeeman effect: $\frac{a_3}{a_1}=|\frac{\rm{cos}(3\pi S)}{\rm{cos}(\pi S)}|$, if $S = n \pm \frac{1}{6}$ where n is an integer, the amplitude of 3F$_\alpha$ vanishes but not the fourth and fifth harmonic \cite{QuantumOscillaitonsSM}.

\section{F$_\gamma$ and F$_\delta$ do not come from Zeeman effect}

We carried out the FFT transformation of the oscillation part on the magnetroresistivity by using different magnetic region (3 T -7 T, 7 T -14 T and 3 T -14 T). The results show that the value of F$_\gamma$ and F$_\delta$ do not shift in different magnetic field region, indicating F$_\gamma$ and F$_\delta$ are two independent bands not from the Zeeman effect of one band.
\begin{figure}
\includegraphics[width=9cm]{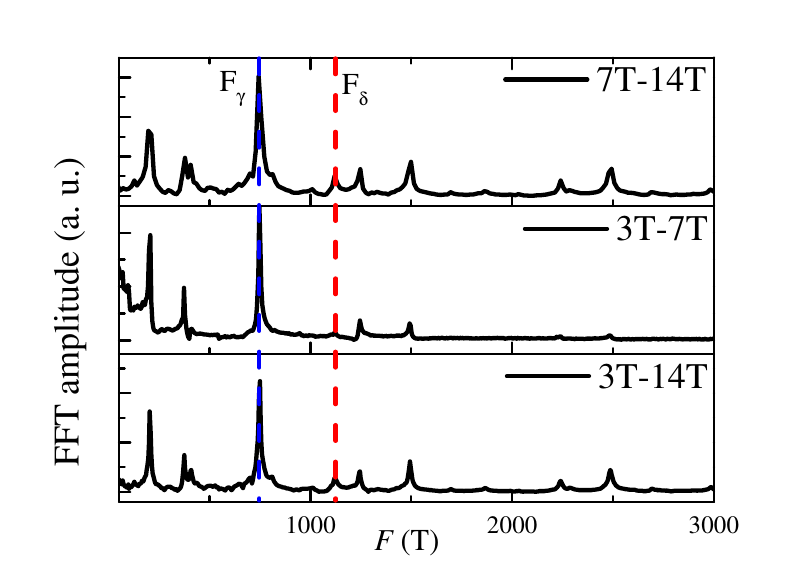}
\caption{The FFT spectrum from different magnetic region (3 T -7 T, 7 T -14 T and 3 T -14 T).}
\end{figure}
\section{Carriers' densities and mobilities fitted by magnetoconductivity and Hall resistivity}
\begin{figure}
\includegraphics[width=9cm]{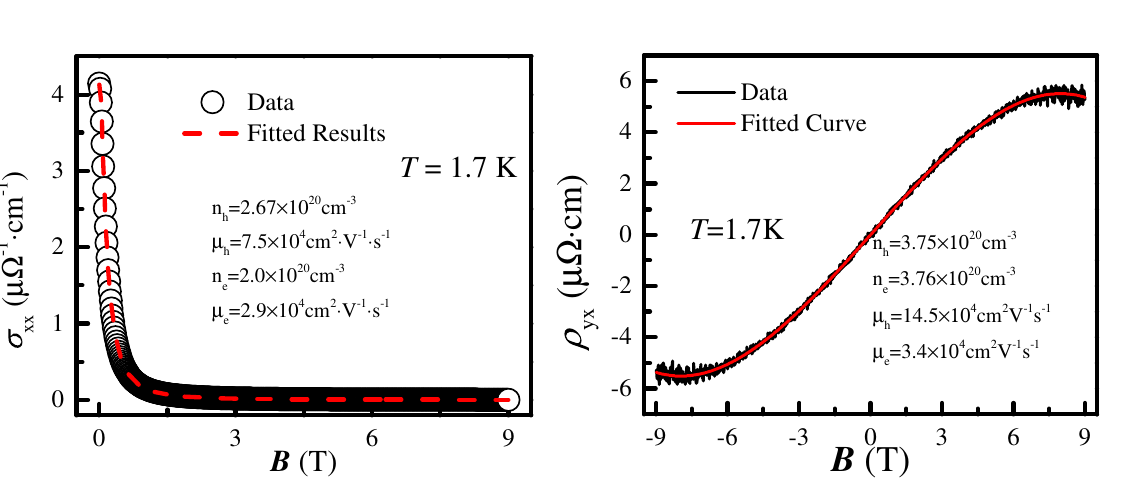}
\caption{Carriers' densities and mobilities fitted by magnetoconductivity (Eq. S1) and Hall resistivity (Eq. S2) .}
\end{figure}

Beside fitted the formula of Hall conductivity used in main text, we also fitted the carriers' densities and mobilities from magnetoconductivity and Hall resistivity using the formulae
\begin{small}
\begin{equation}\label{s}
\sigma_{xx}=\frac{n_e\mu_ee}{1+(\mu_eB)^2}+\frac{n_h\mu_he}{1+(\mu_hB)^2}
\end{equation}
\end{small}
\begin{small}
\begin{equation}\label{s}
\rho_{xy}(B)=\frac{B}{e}\frac{(n_h\mu_h^2-n_e\mu_e^2)+(n_h-n_e)\mu_h^2\mu_e^2B^2}{(n_h\mu_h-n_e\mu_e)^2+(n_h-n_e)\mu_h^2\mu_e^2B^2}
\end{equation}
\end{small} respectively
Here, $n_e$ (or $n_h$) is the carrier density of electrons (or holes), $\mu_e$ ($\mu_h$) is the mobility of electrons (holes). The results are shown in the Table S1:

\begin{table*}[!htb]
\begin{tabular}{| p{3cm} | p{3cm} | p{3cm} | p{3cm} | p{3cm} |}
\hline
 Fitting formulae & $n_e$ (cm$^{-3}$) & $n_h$ (cm$^{-3}$)  & $\mu_e$ (cm$^{-2}$V$^{-1}$s$^{-1}$) & $\mu_h$ (cm$^{-2}$V$^{-1}$s$^{-1}$)  \\
 \hline
 $\sigma_{yx}$(main text)& 1.32$\times$10$^{20}$ &  1.33$\times$10$^{20}$ & 2.8$\times$10$^{4}$ & 15.8$\times$10$^{4}$  \\
 \hline
  $\sigma_{xx}$(Eq.S1)& 2.0$\times$10$^{20}$ &  2.67$\times$10$^{20}$ & 2.9$\times$10$^{4}$ & 7.5$\times$10$^{4}$  \\
 \hline
   $\rho_{yx}$(Eq.S2)& 3.76$\times$10$^{20}$ &  3.75$\times$10$^{20}$ & 3.4$\times$10$^{4}$ & 14.5$\times$10$^{4}$  \\
 \hline
\end{tabular}
\caption{The fitting results for three two-band methods which show consistence of compensation in each fitting method.}
\label{table2}
\end{table*}

We can see that the carrier densities and mobilities obtained by fitting $\sigma_{xy}$, $\sigma_{xx}$ and $\rho_{yx}$ are different but have same order of magnitude. We attribute this difference between different methods to the limitation of the two-carrier fitting method, which is common to fit a multi-band system. \cite {ChenSM, HuangSM}


\begin{thebibliography}{99}

%\bibitem{Bogdanov1989} A. N. Bogdanov and D. A. Yablonskii. Thermodynamically stable ``vortices'' in magnetically ordered crystals. The mixed state of magnets. Sov. Phys. JETP \textbf{68}, 101 (1989).
%\bibitem{Kossler2006} U. K. R\"{o}$\beta$ler, A. N. Bogdanov and C. Pfleiderer.  Spontaneous skyrmion ground states in magnetic metals. Nature \textbf{442}, 797 (2006).
%\bibitem{Suzuki2016} M.-T. Suzuki, T. Koretsune, M. Ochi and R. Arita, arXiv:1611.06042 (2016).

\bibitem{Mangez} J. H. Mangez, J. P. Issi, and J. Heremans, Phys. Rev. B \textbf{14}.10 : 4381 (1976).
\bibitem{Yntema} G. B. Yntema,  Phys. Rev. \textbf{91}, 1388(1953).
\bibitem{Soule} D. E. Soule, Phys. Rev. \textbf{112}, 698(1958).
\bibitem{Mun}E. Mun, H. Ko, G. J. Miller, G. D. Samolyuk, S. L. Bud'ko, and P. C. Canfield, Phys. Rev. B \textbf{85} 035135 (2012).
\bibitem{Takatsu} H. Takatsu, J. J. Ishikawa, S. Yonezawa, H. Yoshino, T. Shishidou, T. Oguchi, K. Murata, and Y. Maeno, Phys. Rev. Lett. \textbf{111} 056601 (2013).
\bibitem{Ali2014}M. N. Ali, J. Xiong, S. Flynn, J. Tao, Q. D. Gibson, L. M. Schoop, T. Liang, N. Haldolaarachchige, M. Hirschberger, N. P. Ong and R. J. Cava, Nature \textbf{514}, 205 (2014).
\bibitem{Cd3As2MR2014} T. Liang, Q. Gibson, M. N. Ali, M. Liu, R. J. Cava, and N. P. Ong, Nat. Mater. \textbf{14}, 280 (2015).
\bibitem{WP2MR2015}N. Kumar Y. Sun, N. Xu, K. Manna, M. Yao, V. S\"{u}ss, I. Leermakers, O. Young, T. F\"{o}rster, M. Schmidt, H. Borrmann, B. Yan, U. Zeitler, M. Shi, C. Felser and C. Shekhar, Nat. Commun. \textbf{8}, 1642 (2017).
\bibitem{LaSbMR2015A}F. F. Tafti, Q. D. Gibson, S. K. Kushwaha, N. Haldolaarachchige, and R. J. Cava, Nat. Phys. \textbf{12}, 272 (2015).
\bibitem{TaAs2MR2016} Y.-Y. Wang, Q.-H. Yu, P.-J. Guo, K. Liu, and T.-L. Xia, Phys. Rev. B \textbf{94}, 041103 (2016).
\bibitem{Xu2017}J. Xu, N. J. Ghimire, J. S. Jiang, Z. L. Xiao, A. S. Botana, Y. L. Wang, Y. Hao, J. E. Pearson, and W. K. Kwok, Phys. Rev. B \textbf{96}, 075159 (2017).
\bibitem{NbPMR2015}C. Shekhar, A. K. Nayak, Y. Sun, M. Schmidt, M. Nicklas, I. Leermakers, U. Zeitler, Y. Skourski, J. Wosnitza, Z. Liu, Y. Chen, W. Schnelle, H. Borrmann, Y. Grin, C. Felser and B. Yan, Nat. Phys. \textbf{11}, 645 (2015).
\bibitem{Zhao2017As}	L. Zhao Q. Xu, X. Wang, J. He, J. Li, H. Yang, Y. Long, D. Chen, H. Liang, C. Li, M. Xue, J. Li, Z. Ren, L. Lu, H. Weng, Z. Fang, X. Dai, and G. Chen, Phys. Rev. B \textbf{95}, 115119 (2017).
\bibitem{NbSb2MR2014} K. Wang, D. Graf, L. Li, L. Wang, and C. Petrovic, Sci. Rep. \textbf{4}, 7328 (2014)
\bibitem{Gao2017PtBi2}	W. Gao,  N. Hao, F. -W. Zheng, W. Ning, M. Wu, X. Zhu, G. Zheng, J. Zhang, J. Lu, H. Zhang, C. Xi, J. Yang, H. Du, P. Zhang, Y. Zhang, and M. Tian, Phys. Rev. Lett. \textbf{118}, 256601 (2017).
\bibitem{BismuthMR2009} B. Fauqu\'{e}, B. Vignolle, C. Proust, J.-P. Issi and K. Behnia , New Journal of Physics \textbf{11}, 113012 (2009).
\bibitem{graphiteMR2017} B. Fauqu\'{e}, and K. Behnia, in Basic Physics of Functionalized Graphite (ed. Esquinazi, P. D.) Ch. 4 (Springer, 2016).
\bibitem{Zhu2015}Z. Zhu, X. Lin, J. Liu, B. Fauqu\'{e}, Q. Tao, C. L. Yang, Y. G. Shi, and K. Behnia, Phys. Rev. Lett. \textbf{114}, 176601(2015).
\bibitem{ZengLaSb2016}L. -K. Zeng, R. Lou, D.-S. Wu, Q. N. Xu, P.-J. Guo, L.-Y. Kong, Y.-G. Zhong, J.-Z. Ma, B.-B. Fu, P. Richard, P. Wang, G. T. Liu, L. Lu, Y.-B. Huang, C. Fang, S.-S. Sun, Q. Wang, L. Wang, Y.-G. Shi, H. M. Weng, H.-C. Lei, K. Liu, S.-C. Wang, T. Qian, J.-L. Luo, and H. Ding, Phys. Rev. Lett. \textbf{117}, 127204 (2016).
\bibitem{Guo2016}P.-J. Guo, H.-C. Yang, B.-J. Zhang, K. Liu, and Z.-Y. Lu, Phys. Rev. B \textbf{93}, 235142 (2016).
\bibitem{LaSb2016Niu}X. H. Niu, D. F. Xu, Y. H. Bai, Q. Song, X. P. Shen, B. P. Xie, Z. Sun, Y. B. Huang, D. C. Peets, and D. L. Feng, Phys. Rev. B \textbf{94}, 165163 (2016).
\bibitem{graphiteMIT2001}D. V. Khveshchenko, Phys. Rev. Lett. 87, 206401 (2001).
\bibitem{graphiteMIT2001b}	Y. Kopelevich, J. C. M. Pantoja, R. R. da Silva, and S. Moehlecke, Phys. Rev. B \textbf{73}, 165128 (2006).
\bibitem{Gibson2015}Q. D. Gibson, L. M. Schoop, L. Muechler, L. S. Xie, M. Hirschberger, N. P. Ong, R. Car, and R. J. Cava, Phys. Rev. B \textbf{91}, 205128 (2015).
\bibitem{Thirupathaiah2018}	S. Thirupathaiah, Y. Kushnirenko, E. Haubold, A. V. Fedorov, E. D. L. Rienks, T. K. Kim, A. N. Yaresko, C. G. F. Blum, S. Aswartham, B. B\"{u}chner, and S. V. Borisenko, Phys. Rev. B \textbf{97}, 035133 (2018).
\bibitem{Yao2016}	Q. Yao, Y. P. Du, X. J. Yang, Y. Zheng, D. F. Xu, X. H. Niu, X. P. Shen, H. F. Yang, P. Dudin, T. K. Kim, M. Hoesch, I. Vobornik, Z.-A. Xu, X. G. Wan, D. L. Feng, and D. W. Shen, Phys. Rev. B \textbf{94} 235140(2016).
\bibitem{XuXF2016}	C. Q. Xu, X. Z. Xing, X. F. Xu, B. Li, B. Chen, L. Q. Che, X. Lu, J. H. Dai, and Z. X. Shi, Phys. Rev. B \textbf{94} 165119(2016).
\bibitem{YangXJ2016}	X. Yang, H. Bai, Z. Wang, Y. Li, Q. Chen, J. Chen, Y. Li, C. Feng, Y. Zheng, and Z.-a. Xu, Appl. Phys. Lett. \textbf{108}, 252401(2016).
\bibitem{Collaudin2015}A. Collaudin, B. Fauqu\'{e}, Y. Fuseya, W. Kang, and K. Behnia, Phys Rev X \textbf{5}, 021022 (2015).
\bibitem{QuantumOscillaitons}	 D. Shoenberg,  Magnetic oscillations in metals, Cambridge University Press. (2009).
\bibitem{SM}See Supplemental Material for the extraction of effective mass from dHvA, magnetoresistance up to 80 T, the quantum oscillations of Hall resistivity, the amplitude of 3F¦Á, discussion on F¦Ã and F¦Ä do not come from Zeeman effect of one band and carriers' densities and mobilities fitted by magnetoconductivity and Hall resistivity.
\bibitem{Sb2018} B. Fauqu\'{e}, X. Yang, W. Tabis, M. Shen, Z. Zhu, C. Proust, Y. Fuseya, and K. Behnia, arXiv:1803.00931
\bibitem{SbDensity}L. R. Windmiller, Phys. Rev. \textbf{149}, 472 (1966)
\bibitem{ShallowFS}K. Behnia J. Phys.: Condens. Matter \textbf{27} 375501 (2015)
\bibitem{He2016} J. He, C. Zhang, N. J. Ghimire, T. Liang, C. Jia, J. Jiang, S. Tang, S. Chen, Yu He, S.-K. Mo, C. C. Hwang, M. Hashimoto, D. H. Lu, B. Moritz, T. P. Devereaux, Y. L. Chen, J. F. Mitchell, and Z.-X. Shen, Phys. Rev. Lett. \textbf{117}, 267201 (2016)
\bibitem{Pavlosiuk2016} O. Pavlosiuk, P, Swatek and P, Wi\'{s}niewski, Sci. Rep. \textbf{6}, 38691 (2016)
\bibitem{bismuthF}Z. Zhu, B. Fauqu\'{e}, K. Behnia, Y. Fuseya, arXiv:1801.07098
%Z. Zhu, R. D. McDonald, A. Shekhter, B. J. Ramshaw, K. A. Modic, F. F. Balakirev, and N. Harrison, Sci. Rep. 7, 1733 (2017).
\bibitem{YSbanistropy}O. Pavlosiuk, P. Swatek, and P. Wi\'{s}niewski, Sci. Rep. \textbf{6}, 38691 (2016).
\bibitem{YSbmass}Q. H. Yu, Y.-Y. Wang, R. Lou, P. -J. Guo, S. Xu, K. Liu, S. Wang, and T. -L. Xia, Europhys. Lett. \textbf{119}, 17002 (2017)
\bibitem{WTe2mass}F. -X. Xiang, M. Veldhorst, S. -X. Dou, and X. -L. Wang, Europhys. Lett. \textbf{112}, 37009 (2015)
%\bibitem{TaAsMR2015}X. Huang et al., Phys Rev X 5, 031023 (2015).
%\bibitem{TaAsARPES2015}B. Q. Lv et al., Nat Phys 11, 724 (2015)
%\bibitem{spin1965a}	M. Fowler, R. E. Prange,  Physics, 1, 315 (1965)
%\bibitem{LK}I.M. Lifshitz, A.M. Kosevich, Zh. Eksp. Teor. Fiz. 29 (1955) 730.
\end{thebibliography}

\begin{thebibliography}{99}

\bibitem{QuantumOscillaitonsSM}	 D. Shoenberg,  Magnetic oscillations in metals, Cambridge University Press. (2009).
\bibitem{ChenSM}D. Chen, L. X. Zhao, J. B. He, H. Liang, S. Zhang, C. H. Li, L. Shan, S. C. Wang, Z. A. Ren, C. Ren, and G. F. Chen Phys. Rev. B \textbf{94}, 174411 (2016)
\bibitem{HuangSM}Xiaochun Huang, Lingxiao Zhao, Yujia Long, Peipei Wang, Dong Chen, Zhanhai Yang, Hui Liang, Mianqi Xue, Hongming Weng, Zhong Fang, Xi Dai, and Genfu Chen Phys. Rev. X 5, 031023 (2015)
\end{thebibliography}
\end{document}